\documentclass[%
reprint,
superscriptaddress,
 amsmath,amssymb,
 aps,
prl,
floatfix,
usenames,dvipsnames
]{revtex4-2}

\usepackage{xcolor}
\usepackage{graphicx}
\usepackage{dcolumn}
\usepackage{bm}
\usepackage{physics}
\usepackage{mathtools}
\usepackage{enumitem}
\usepackage{subeqnarray}

\newcommand{\chit}{\chi^{(2)}}

\renewcommand{\exp}[1]{\,e^{#1}}

\newcommand{\ai}{\hat{a}_i}
\newcommand{\aid}{\hat{a}^\dagger_i}

\newcommand{\asd}{\hat{a}^\dagger_s}
\newcommand{\sg}{\hat{\sigma}}
\newcommand{\sgd}{\hat{\sigma}^\dagger}

\begin{document}

\preprint{APS/123-QED}

\title{A hybrid source of quantum light for generation of frequency tunable Fock states
}

\author{Aleksa Krsti\' c}
\email{aleksa.krstic@uni-jena.de}
\affiliation{
 Institute of Applied Physics, Abbe Center of Photonics, Friedrich-Schiller University Jena, Albert-Einstein-Straße 15, 07745 Jena, Germany}

\author{Priyanshu Tiwari} 
\affiliation{
 Institute of Applied Physics, Abbe Center of Photonics, Friedrich-Schiller University Jena, Albert-Einstein-Straße 15, 07745 Jena, Germany}

 \author{Florian Höhe}
 \affiliation{Institute for Complex Quantum Systems and IQST, University of Ulm, Albert-Einstein-Allee 11, 89069 Ulm, Germany}
 
\author{Frank Setzpfandt}
\affiliation{
 Institute of Applied Physics, Abbe Center of Photonics, Friedrich-Schiller University Jena, Albert-Einstein-Straße 15, 07745 Jena, Germany}
\affiliation{Fraunhofer Institute for Applied Optics and Precision Engineering, Albert-Einstein-Straße 7, 07745 Jena, Germany}

\author{Ulf Peschel}
\affiliation{Institute of Condensed Matter Theory and Solid State Optics, Friedrich-Schiller University Jena, Max-Wien-Platz 1, 07743 Jena, Germany}

\author{Joachim Ankerhold}
\affiliation{Institute for Complex Quantum Systems and IQST, University of Ulm, Albert-Einstein-Allee 11, 89069 Ulm, Germany}
 
\author{Sina Saravi}
\affiliation{
 Institute of Applied Physics, Abbe Center of Photonics, Friedrich-Schiller University Jena, Albert-Einstein-Straße 15, 07745 Jena, Germany}%

\date{\today}

\begin{abstract}
We propose a scheme for quantum-light generation in a nonlinear cavity hybridized with a 2-level system and theoretically show that, when excited by a series of controlled pump pulses, the hybrid source generates Fock states with high probabilities. E.g., 1- and 2-photon states can be generated near-on-demand, and Fock states with up to $7$ photons with a probability above $50\%$. The tailorable nature of the nonlinear cavity allows for generating Fock states with arbitrary frequencies, even with a fixed 2-level system, creating fundamentally new opportunities in all areas of quantum technologies.
\end{abstract}

\maketitle

\noindent
\textit{Introduction.}
To support the realization of optical quantum technologies \cite{flamini2018photonic, weedbrook2012gaussian, brod2019CV-q-computation,perarnau2020q-metrology}, development of efficient and tailorable sources of quantum light, such as squeezed light and Fock states, is of major interest.
While squeezed light can reliably be generated using nonlinear parametric processes \cite{schnabel2016squeeze, barsotti2018squeezed}, generation of optical Fock states with photon numbers $n\geq 2$ is notoriously difficult in practice and is an ongoing topic of research. The majority of proposals and implementations rely on one of two general approaches: using atomic or atom-like solid-state quantum systems \cite{solano2020deterministicFock,PhysRevA.67.043818,PhysRevLett.115.163603,PhysRevResearch.2.033489,PhysRevA.67.043818}, with successful implementations mainly based on temporal demultiplexing of single photons from a single quantum dot \cite{lenzini2017active,PhysRevLett.123.250503,Hansen:23}, or using nonlinear parametric sources, where Fock states can be heralded by conditional measurement of squeezed-light states
\cite{waks2006generation,cooper2013experimental,tiedau2019SPDC_Fock}.
Yet both approaches have their limitations. Although capable of deterministic generation, atomic and solid-state systems have very limited tunabilities \cite{lee2020integrated}, especially in their spectra, which result in states with limited tailorability in spectral/modal properties.
In contrast, nonlinear sources are widely tunable in spectral, modal, and polarization degrees of freedom, specially in nanostructured platforms \cite{wang2021integrated}. Yet, they only generate Fock states probabilistically, with decreasing probability for higher photon-number states \cite{tiedau2019SPDC_Fock}.

Motivated by these complementary advantages and disadvantages of nonlinear and atom-like systems, we aim for a hybrid system combining the two, that exploits both their advantages towards creating an ideal source of Fock states, one that is on-demand and tunable.
Such hybrid systems have attracted attention recently, with proposals for enhancing atom-cavity interaction \cite{qin2018exponentially,PhysRevLett.120.093602}, single-photon generation with enhanced purity \cite{saravi2017AMSPDC}, generating equally weighted superposition of Fock states \cite{ping2022hybridSuperpositions}, and generating Schrödinger cat states \cite{chen2021shortcuts}.

In this Letter, we propose a hybrid source, consisting of a nonlinear cavity and a two-level system (2LS), for generation of Fock states with enhanced probabilities and arbitrary frequencies. To this end, we use a sequence of optical pump pulses for the nonlinear cavity, with controlled amplitudes, phases, and temporal delays, and leverage the complex interplay between squeezed-light generation and its interaction with a 2LS. Importantly, the properties of the generated Fock states, specially their frequency, can be adjusted by tuning the parameters of the nonlinear process. We also study this system's performance in the presence of cavity losses and dephasing of the 2LS. 

\begin{figure*}[htpb!]
    \includegraphics[width=\textwidth]{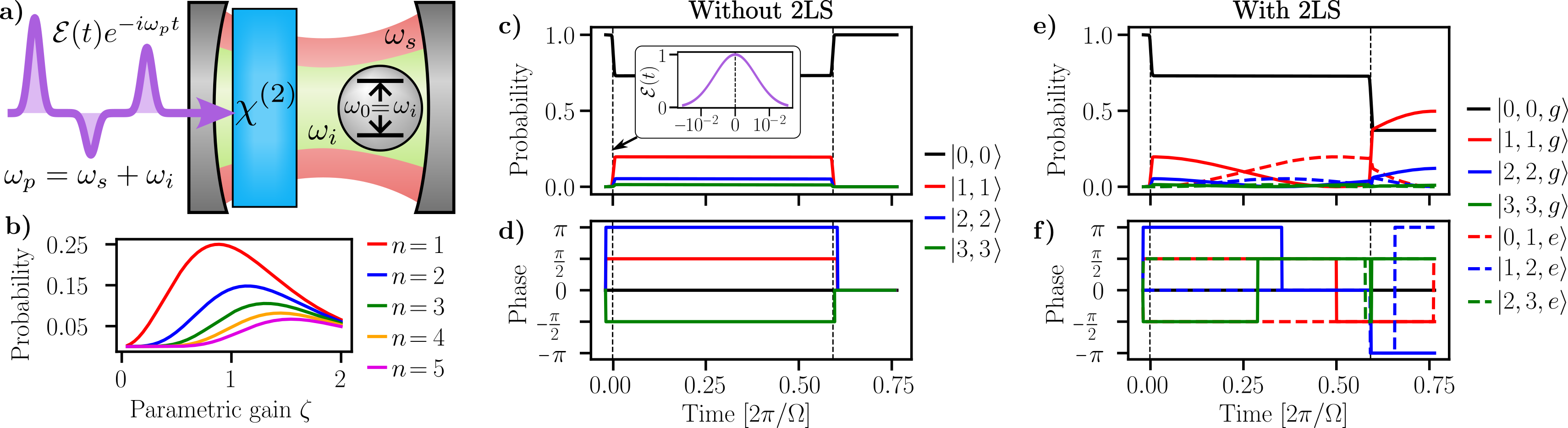}
    \caption{\textbf{(a)} Schematic representation of the hybrid source. A sequence of pump pulses with distinct temporal separations, amplitudes, and phases are incident on a nonlinear cavity, to generate a two-mode squeezed vacuum (TMSV) in the signal and idler modes. Simultaneously, the idler photons undergo resonant Rabi oscillations with a 2LS. \textbf{(b)} Probabilities of obtaining the five lowest order $\ket{n,n}$ states as a function of a real-valued parametric gain $r$, in a conventional TMSV source without the 2LS. \textbf{(c,d)} Temporal evolution of the probabilities (c) and phases of probability amplitudes (w.r.t. the $\ket{0,0}$ state) (d) for the three lowest order $\ket{n,n}$ in a configuration with two pump pulses and no 2LS. The pulses have the parametric gains $r_1=-r_2\approx0.58$ and a temporal separation of $T_1=0.59\frac{2\pi}{\Omega}$. Vertical black dashed lines indicate the arrival time of each pump pulse. As the pulses are too short to be discernible at these time scales, a plot of a pump pulse envelope is shown in the inset of (c); \textbf{(e,f)} Temporal evolution of a few lowest-order composite states $\ket{n,n,g}$ and $\ket{n-1,n,e}$ (phases w.r.t. the $\ket{0,0,g}$ state), in the nonlinear cavity with the 2LS, for the same pumping configuration as in (c,d).}
    \label{fig1}
\end{figure*}

\noindent\textit{Basic principles.} 
The hybrid source is depicted schematically in Fig.~\ref{fig1}(a), a nonlinear crystal with second-order susceptibility $\chit$ and a 2LS, are embedded in an optical cavity.
The nonlinear cavity, without the 2LS, when pumped by an optical field of frequency $\omega_p=\omega_s+\omega_i$, can generate two-mode squeezed vacuum (TMSV) into two separate resonance modes of the cavity, referred to as \textit{signal} and \textit{idler} modes, of frequencies $\omega_s$ and $\omega_i$, respectively. The 2LS has the transition frequency $\omega_0$, which is resonant with the idler mode, i.e., $\omega_i=\omega_0$.
The TMSV state generated by the squeezing operator $\exp{-i \left(r \aid\asd+\ \text{H.c.}\right)}$ can be written as a superposition of perfectly photon-number correlated states as $\ket{\text{TMSV}}=\cosh^{-1}(\zeta)\sum_{n=0}^\infty (i\exp{i\phi} \tanh(\zeta))^n\ket{n,n}$ \cite{caves1985squeeze,eckstein2011TMSV}. Here, $\aid$ and $\asd$ are the creation operators for the idler and signal modes, respectively. We adopt the convention that a state of the form $\ket{m,n}$ contains $m$ idler, and $n$ signal photons and also refer to states where $n=m$ as \textit{multi-pair} states. Additionally, $\zeta$ and $\phi$ are the magnitude and phase, respectively, of the complex parametric gain $r=\zeta\exp{i\phi}$, determined by the properties of the nonlinear cavity and the pump field \cite{caves1985squeeze}. 

In Fig.~\ref{fig1}(b), we show the probability of finding the first few multi-pair states in a TMSV state, as a function of the parametric gain. We observe that, as the parametric gain increases, these probabilities rise to a maximum value and then decrease. This is a consequence of lower-order $\ket{n,n}$ states "seeding" the generation of the higher-order ones. This property of squeezed-light sources sets a fundamental limit on the probability of generating any given multi-pair state. 
In principle, a combination of heralding and multiplexing approaches can be used to enhance the generation probability of a specific Fock state. Yet, this demands enormous resources in practice, even for the simplest case of realizing an on-demand source of single photons \cite{silberhorn2020Multiplexing,saravi2021LNOI}.  
 
\noindent\textit{Hybrid dynamics.} The hybrid system dynamics can be described in the interaction picture, with the Hamiltonian
$\hat{H}(t)=\hbar \Gamma \mathcal{E}(t)\aid\asd+i \hbar \frac{\Omega}{2}\ai\sgd +\ \text{H.c.}$ \cite{scully1997QO,jaynes1963JCmodel}. 
Here, $\Gamma$ is a nonlinear coupling constant, representing the nonlinear efficiency of the cavity. $\Omega$ is the single-photon Rabi frequency, and $\sg$ is the atomic lowering operator for the 2LS.
$\mathcal{E}(t)$ is the temporal envelope of the pump field, which is made of a sequence of temporally distinct pulses.
Each pump pulse is enumerated by the number $j$, and is characterized by a parametric gain
$r_j=\zeta_j\exp{i\phi_j}=\Gamma\int_j\dd{t}\mathcal{E}(t)$, where the integration extends over the duration of the $j$-th pulse.
Without loss of generality, we use Gaussian pump pulses with temporal widths fixed to $\approx5\times10^{-3}\frac{2\pi}{\Omega}$, much shorter than the temporal period of few-photon Rabi oscillations between the idler photons and the 2LS, which allows us to isolate the dynamics of nonlinear generation from the Rabi oscillations. Essentially, with every shot of a pump pulse, the nonlinear Hamiltonian induces a quick change in the system's state, followed by much slower Rabi oscillations.
The details of the analytical derivations and numerical methods used \cite{johansson2012qutip} are outlined in the Supplementary \cite{supp}.
Finally, to better understand the physics of the process, we neglect dissipative processes (e.g., cavity loss) for the time being and discuss them in a later section.

To understand the hybrid dynamics, an example with a two-pulse pump sequence is shown in Fig.~\ref{fig1}(c-f). We plot the temporal evolution of the first few multi-pair states, where we show their probabilities and their relative phases with respect to the $\ket{0,0}$ state.
In Fig.~\ref{fig1}(c,d), we study the case of the nonlinear cavity without the 2LS, essentially a TMSV source. 
We use two pump pulses of equal gain magnitudes, but with a relative phase difference of $\pi$, with $r_1=-r_2\approx0.58$.
The sharp changes in all the plots correspond to the points in time where a pump pulse is incident on the nonlinear cavity.
As we see in this case, these two pulses cancel each others parametric gains, resulting in a perfect destructive nonlinear interference \cite{chekhova2016NonlInterf}, creating the TMSV with the first pulse and going back to the vacuum state with the second pulse.

In Figs.~\ref{fig1}(e,f), we show the results for the same two-pulse configuration as in Figs.~\ref{fig1}(c,d), but now in the presence of the 2LS in the nonlinear cavity.
Firstly, we see that for each individual multi-pair state generated by the first pump pulse, the hybrid system experiences Rabi oscillations between composite states $\ket{n,n,g}$ and $\ket{n-1,n,e}$, where $g$ ($e$) indicates the ground (excited) state of the 2LS.
The frequency of each Rabi oscillation is proportional to $\sqrt{n}$ \cite{jaynes1963JCmodel,scully1997QO}.
This scaling of the Rabi frequency with photon number results in a complex set of relative phases and amplitudes between individual multi-pair states by the time the second pump pulse arrives. Consequently, this changes the nature of nonlinear interferences and results in a final state that differs significantly from the vacuum state obtained in the case without the 2LS. We discuss this mechanism in detail in the Supplementary \cite{supp}.
Interestingly, we can see in Fig.~\ref{fig1}(e), that the probability of generating a single pair of signal and idler photons reaches a value close to $0.5$, which is twice the value than can be achieved in a conventional TMSV source, as was seen in Fig.~\ref{fig1}(b).

\begin{figure}[t]
	\includegraphics[width=\linewidth]{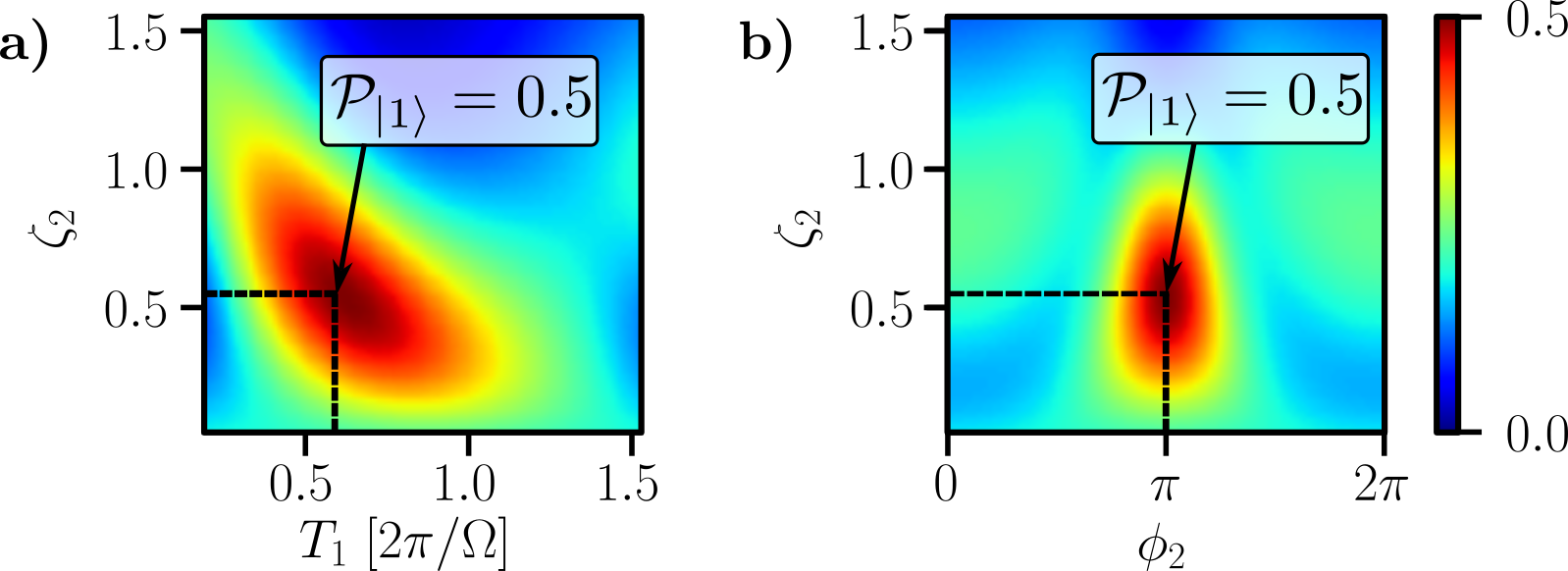}
	\caption{\textbf{(a)} Probability of having a single photon in the signal mode in a two-pulse configuration, as a function of the parametric gain magnitude $\zeta_2$ and delay time $T_1$ of the second pump pulse (fixed relative phase, $\phi_2=\pi$ and $\phi_1=0$); \textbf{(b)} The same probability as a function of the phase and magnitude of $r_2=\zeta_2\exp{i\phi_2}$ (delay time fixed to $T_1=0.61\frac{2\pi}{\Omega}$). In both cases, the gain of the first pulse is $r_1\approx0.58$.}
	\label{fig2}
\end{figure}

In fact, the final state varies significantly with the parametric gain, relative phase, and the delay time between the pump pulses. To see this, in Fig.~\ref{fig2}, we show the probability of finding a single-photon state in the signal mode (calculated by adding the probabilities of the $\ket{1,1,g}$ and $\ket{0,1,e}$ states): (a) as a function of the delay time $T_1$ and gain magnitude $\zeta_2$ of the second pulse, for a fixed relative phase $\phi_2-\phi_1=\pi$; (b) as a function of the phase $\phi_2$ and magnitude $\zeta_2$ of the parametric gain of the second pulse, for a fixed delay time $T_1=0.61\frac{2\pi}{\Omega}$. The other parameters were identical to the configuration in Figs.~\ref{fig1}(e,f). In Fig.~\ref{fig2}(a), we see that the single-photon generation probability attains a maximum of $0.5$ for a delay of $T_1=0.61\frac{2\pi}{\Omega}$ and a parametric gain of $\zeta_2\approx0.55$. In Fig.~\ref{fig2}(b), we see that phases $\phi_2\neq\pi$ always result in a lower single-photon generation probability. 
Further simulations (see Supplementary \cite{supp}) also showed that, within the considered ranges of $\zeta_2$ and $T_1$, the highest single-photon probabilities are always obtained for a relative phase of $\pi$ between two consecutive pump pulses.

\noindent\textit{Three-pulse configuration.} 
While the high-dimensional space of control parameters in a multi-pulse sequence offers many ways to tune desired outputs, even configurations with a limited number of varying parameters are capable of generating high-photon-number Fock states, with generation probabilities far surpassing what is possible in a conventional TMSV source. 
To demonstrate this, we numerically calculate the output for a configuration with three pump pulses \cite{johansson2012qutip}. We use the parametric gains and delay times of the pulses as "control knobs" for tuning the system towards desired final states, namely to maximize the generation probability of different Fock states in the signal mode of the cavity.
We fixed the relative phase between the pump pulses to $\pi$, such that $\phi_1=0$, $\phi_2=\pi$, and $\phi_3=0$. This choice is guided by our study shown in Fig.~\ref{fig2}, to maximize the single-photon probability in the signal mode. This does not necessarily mean that this phase choice is a global optimum for maximizing the generation probability of other Fock states, yet it is a reasonable choice, and mainly an attempt to limit the space of control parameters in our optimization study.

For the parametric gain magnitudes, we consider values of $\zeta\leq15\text{dB}$, which are experimentally achievable in TMSV sources \cite{schnabel2016squeeze}. From this point onward, we express parametric gain magnitudes in units of dB, calculated as $\zeta[\text{dB}]=-10\log_{10}(\exp{-2\zeta})$, for easier comparison with experimental works on TMSV sources. We also restrict the total interaction time to $\leq3\frac{2\pi}{\Omega}$, as the duration of coherent interactions between a quantum emitter and an optical cavity is limited in practice \cite{Lodahl2015}.
In the next section, we will briefly address the feasibility of realizing our optimized pulse sequences, in terms of the required interaction time, with respect to such realistic effects.

\begin{figure}[htpb!]
	\includegraphics[width=0.98\linewidth]{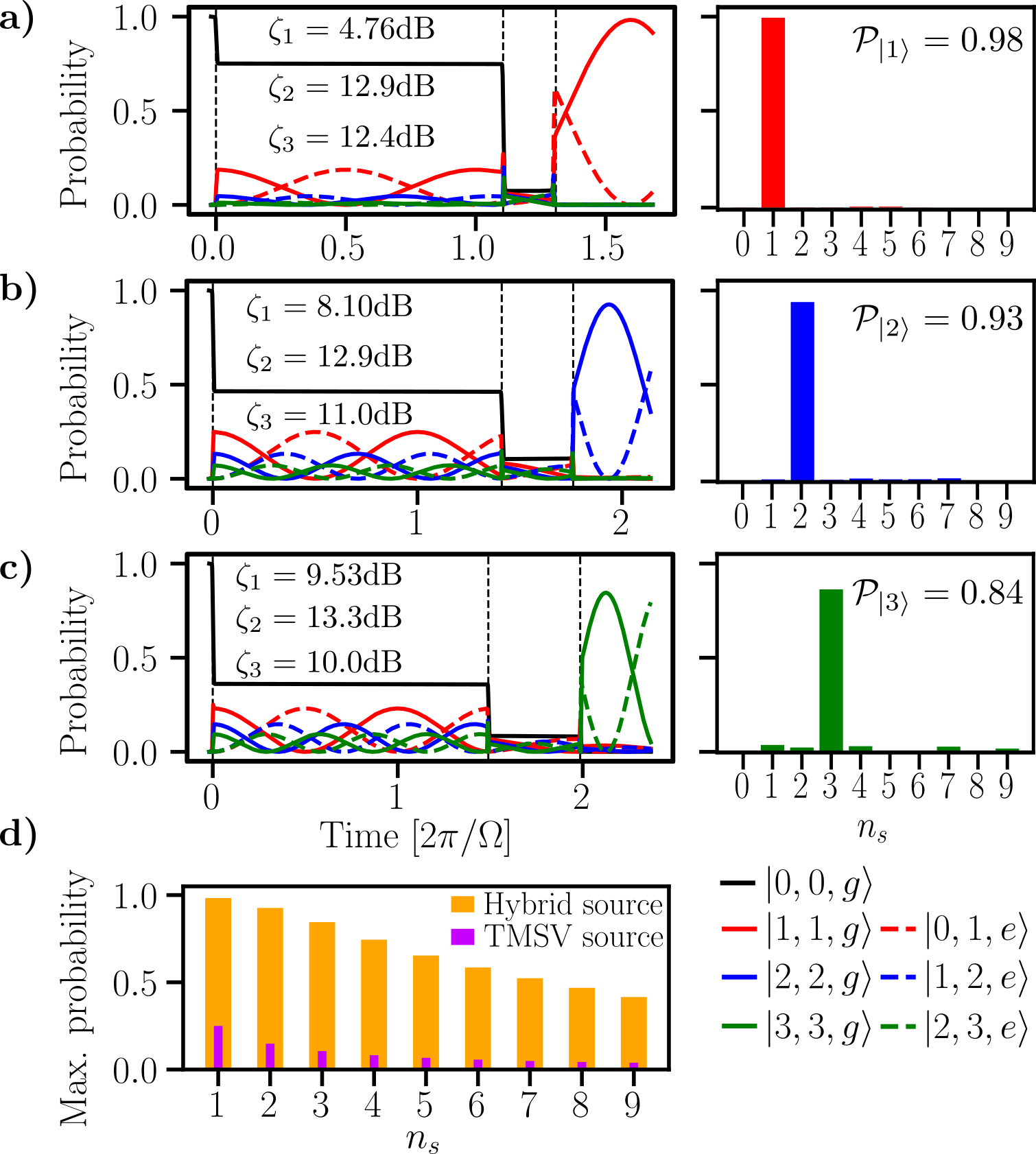}
	\caption{\textbf{(a-c)} Left panels: Temporal evolution of probabilities in a configuration with three pump pulses, optimized for achieving maximum generation probability of (a) $1$, (b) $2$, and (c) $3$ photons in the signal mode. The second pulse has a relative phase of $\pi$ compared to the other two and the parametric gain of each pulse is indicated in each figure as $\zeta_j$. Right panels: The corresponding photon-number probability distributions in the signal mode at the output. \textbf{(d)} The maximum obtained probabilities for generating photon number states $\ket{1}$ to $\ket{9}$ in the signal mode, in the three pulse configuration (yellow bars), and the maximum fundamentally attainable  probabilities for generating the $\ket{n,n}$ states in a conventional TMSV source (purple bars).}
	\label{fig3}
\end{figure}

Under the expressed boundary conditions for our optimization with three pump pulses, we show in Figs.~\ref{fig3}(a,b,c), the pulse configurations that result in the maximum achievable generation probability for the Fock states $\ket{1}$, $\ket{2}$ and $\ket{3}$ in the signal mode, respectively.
The probability of generating an $n$-photon Fock state in the signal mode is calculated from summing up the mutually exclusive probabilities of generating the $\ket{n,n,g}$ and $\ket{n-1,n,e}$ states. As the signal mode does not interact with the 2LS, the Fock state in it can eventually escape a realistic cavity of finite quality factor for further utilization.
The left panels in Figs.~\ref{fig3}(a-c) show the temporal evolution of the probabilities, while the right panels show the final distribution of the probabilities for the signal-mode Fock states, where we find very high probabilities for generation of these distinct Fock states. Especially Figs.~\ref{fig3}(a,b), show a near-on-demand generation of single- and two-photon Fock states, with $0.98$ and $0.93$ probabilities, respectively. 
In Fig.~\ref{fig3}(d), we show the maximum probabilities for generation of Fock states with photon numbers up to $n_s=9$, that can be obtained in our three-pulse optimization. With it, we also show the corresponding maximum probabilities attainable using a conventional TMSV source for comparison, which clearly demonstrates the enhancement achieved in the hybrid system. As an example, the probability for generating the $n_s=9$ state is enhanced more than 10-fold compared to the TMSV source.
The exact pulse sequence parameters for attaining each of the Fock states shown in Fig.~\ref{fig3}(d) are given in the Supplementary \cite{supp}.

\begin{figure}[b]
	\includegraphics[width=0.95\linewidth]{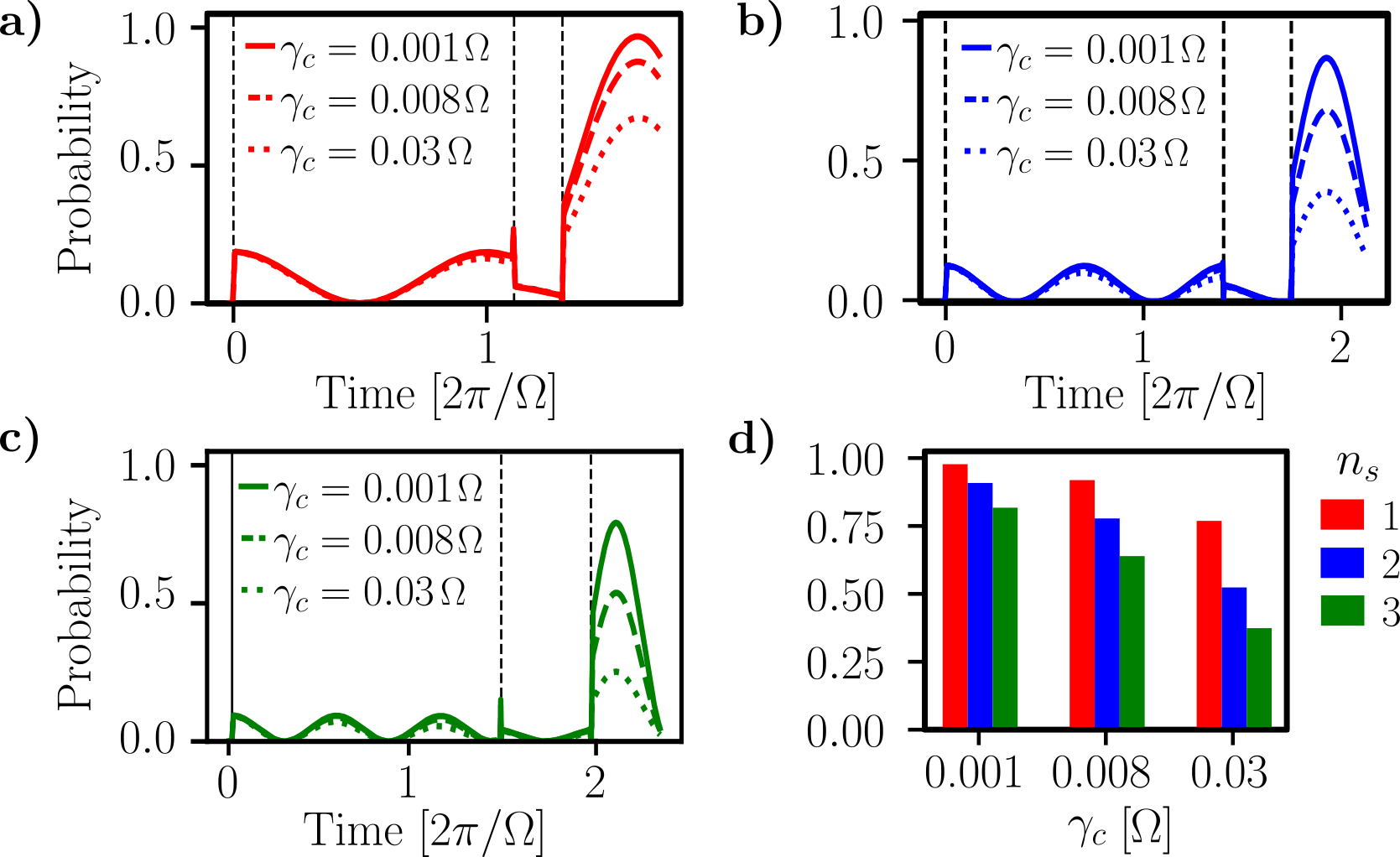}
	\caption{\textbf{(a-c)} Temporal evolution of the $\ket{1,1,g}$ (red), $\ket{2,2,g}$ (blue), and $\ket{3,3,g}$ (green) states in the presence of different amounts of cavity losses, quantified by $\gamma_c$, the decay rate for both the signal and idler modes. The pump configuration for each of the plots is identical to the corresponding optimized lossless cases shown in Fig.~\ref{fig3}; \textbf{(d)} The probabilities of detecting $n_s=1,2,3$ photons in the signal mode immediately after the third pump pulse.}
	\label{fig4}
\end{figure}

\noindent\textit{Effects of loss.} 
The obtained results up to now are only strictly valid in an idealized system, with no cavity losses, $100\%$ coupling of the 2LS radiation into the cavity mode, and no internal dephasing of the 2LS, which are not achievable in practice \cite{Lodahl2015}. 

To take the first step into investigating our scheme under non-ideal conditions, we consider the effects of the cavity having a finite quality factor, as well as pure dephasing within the 2LS. We do this by solving the Lindblad master equation
$\partial_t\hat{\rho}=\frac{1}{i\hbar}[\hat{H},\rho]-\sum_{j=i,s,d}\frac{1}{2}(\hat{c}_j^{\dagger}\hat{c}_j\rho+\rho\hat{c}_j^{\dagger}\hat{c}_j-2\hat{c}_j\rho\hat{c}_j^{\dagger})$,
where $\hat{c}_j=\sqrt{\gamma_c}\hat{a}_j$ for $j=\{i,s\}$ are the jump operators for the cavity loss with $\gamma_\text{c}$ as the cavity decay rate, and $\hat{c}_d=\sqrt{\frac{\gamma_\text{d}}{2}}\hat{\sigma}_z$ is the jump operator for pure 2LS dephasing with the dephasing rate $\gamma_\text{d}$ and $\hat{\sigma}_z=\comm{\hat{\sigma}^\dagger}{\hat{\sigma}}$ \cite{Lodahl2015}. We simulated the non-ideal hybrid system for the same sets of pump parameters as in Fig.~\ref{fig3}(a-c) and, to more easily quantify the influence of each dissipative mechanism, we investigate their effects on the hybrid system separately.

Our simulations show the hybrid system to be quite robust to dephasing for dephasing rates commonly measured for solid-state emitters \cite{press2010ultrafast,dietrich2016gaas,esmann2024solid}, with a dephasing rate of $\gamma_d=0.05\Omega$ resulting in less than $10\%$ reduction in the output probabilities of the optimized configurations. This result, along with our full dephasing study, is given in detail in the supplementary \cite{supp}. Here we focus on cavity losses which have a more prominent effect on the system’s performance. 
This is shown in Fig.~\ref{fig4}, where we plot the temporal evolution of the $\ket{1,1,g}$ [(a)], $\ket{2,2,g}$ [(b)], and $\ket{3,3,g}$ [(c)] states for three values of the decay rate $\gamma_c=\{0.001,0.008,0.03\}\Omega$. As expected, the cavity loss reduces the probabilities of obtaining the target states, with the higher-order states
suffering more due to their higher photon number. 
As a visual summary, the probabilities for obtaining a Fock state with $n_s=1,2,3$ photons in the signal mode immediately after the third pulse are shown in Fig.~\ref{fig4}(d), where each set of bars corresponds to a particular value of $\gamma_c$ and the values shown are obtained by tracing-out the idler and 2LS degrees of freedom.
In the supplementary \cite{supp}, we also present a rough calculation to connect the used $\gamma_c$ values to the Purcell factor of an optical cavity, considering common solid-state-emitter parameters for the 2LS \cite{Lodahl2015, he2015single, lee2020integrated}, where in our estimation, state-of-the-art micro/nanostructured optical cavities can reach these requirements \cite{deotare2009high,PhysRevX.10.041025,PhysRevApplied.18.044066}. We also present a detailed discussion in the supplementary \cite{supp} about the current technological platforms that could be used for realizing such a hybrid system \cite{schnabel2016squeeze,azzini2013stimulated,marty2021photonic,lu2021ultralow,steiner2021ultrabright,orieux2013direct,senellart2017high,ding2016demand,barone2024generation,zhang2021squeezed,johnston2024cavity,senichev2021room,li2024integrated,castelletto2021silicon,elshaari2020hybrid,aghaeimeibodi2018integration,park2024single,nehra2022few,peyskens2019integration}.

\noindent\textit{Discussion and conclusion.}
Our results are obtained by considering only a portion of the full control-parameter space of the hybrid system. Further expanding the search to include more pulses, continuous relative phase variations, higher parametric gains, and frequency detuning between the cavity modes and the 2LS, could lead to further enhancing the generation probabilities, to potentially reach near-deterministic generation of higher Fock states, or even other classes of non-Gaussian states for continuous variable quantum information processing \cite{PRXQuantum.2.030204}.
Moreover, we only investigated loss with configurations optimized under lossless conditions. With an optimization under lossy conditions, one might identify pulse sequences that make the process more robust to losses.

Importantly, the hybrid source inherits the tunability of a TMSV source. Specifically, the frequency spectrum of the output photons in the signal mode can be fully controlled by the energy conservation relation $\omega_s=\omega_p-\omega_0$, only by tuning the frequency spectrum of the pump pulses and the resonances of the nonlinear cavity, even while the 2LS is unchanged. 
This opens the door to fundamentally new opportunities for optical quantum technologies, starting with near-deterministic generation of Fock states for different applications in quantum computing, communication, and sensing, which commonly operate in very different wavelength ranges. Moreover, with this scheme, one could in principle realize many indistinguishable sources of Fock states by using many frequency-distinguishable solid-state emitters \cite{lee2020integrated}.

Finally, although we had photonic systems in mind, the predicted dynamics are independent of the frequency range of operation and could potentially be adapted to other quantum information processing platforms, e.g. for microwave quantum-state generation in superconducting circuits \cite{PhysRevLett.122.186804,PhysRevX.11.031008}, motional quantum-state generation in trapped ions \cite{PhysRevLett.76.1796, PhysRevLett.122.030501}, and generating phonon-photon excitations in optomechanical systems \cite{barzanjeh2022optomechanics}.

\begin{acknowledgments}
This research is supported by funding from the Carl-Zeiss-Stiftung (CZS Center QPhoton), the German Research Foundation (DFG) under the project identifier 398816777-SFB 1375 (NOA), and the German Federal Ministry of Education and Research (BMBF) under the project identifiers 13N14877 (QuantIm4Life) and 13N16108 (PhoQuant).
J. A. and F. H. acknowledge financial support by the German Science Foundation (DFG) through AN336/18-1 and AN336/13-1. U. P. acknowledges support by the 
Thuringian Ministry for Economy, Science, and Digital Society (Quantum Hub Thüringen project AP Qi2.5) and S. S. acknowledges funding by the Nexus program of the Carl-Zeiss-Stiftung (project MetaNN). 
\end{acknowledgments}

\bibliography{bibliography}

\end{document}


\preprint{Supplemental material}

\title{A hybrid source of quantum light for generation of frequency tunable Fock states}

\author{Aleksa Krsti\' c}
\email{aleksa.krstic@uni-jena.de}
\affiliation{
 Institute of Applied Physics, Abbe Center of Photonics, Friedrich-Schiller University Jena, Albert-Einstein-Straße 15, 07745 Jena, Germany}

\author{Priyanshu Tiwari} 
\affiliation{
 Institute of Applied Physics, Abbe Center of Photonics, Friedrich-Schiller University Jena, Albert-Einstein-Straße 15, 07745 Jena, Germany}

 \author{Florian Höhe}
 \affiliation{Institute for Complex Quantum Systems and IQST, University of Ulm, Albert-Einstein-Allee 11, 89069 Ulm, Germany}
 
\author{Frank Setzpfandt}
\affiliation{
 Institute of Applied Physics, Abbe Center of Photonics, Friedrich-Schiller University Jena, Albert-Einstein-Straße 15, 07745 Jena, Germany}
\affiliation{Fraunhofer Institute for Applied Optics and Precision Engineering, Albert-Einstein-Straße 7, 07745 Jena, Germany}

\author{Ulf Peschel}
\affiliation{Institute of Condensed Matter Theory and Solid State Optics, Friedrich-Schiller University Jena, Max-Wien-Platz 1, 07743 Jena, Germany}

\author{Joachim Ankerhold}
\affiliation{Institute for Complex Quantum Systems and IQST, University of Ulm, Albert-Einstein-Allee 11, 89069 Ulm, Germany}
 
\author{Sina Saravi}
\affiliation{
 Institute of Applied Physics, Abbe Center of Photonics, Friedrich-Schiller University Jena, Albert-Einstein-Straße 15, 07745 Jena, Germany}%

\date{\today}

\maketitle
\tableofcontents

\section{The interaction Hamiltonian and physical interpretations of the hybrid dynamics}\label{sec:interaction_etc}
The nonlinear interaction inside the crystal and the interaction of the emitter with the idler mode are described by the total Hamiltonian $\hat{H}(t)=\hat{H}_0+\hat{H}_{\text{NL}}(t)+\hat{H}_{\text{JC}}$. Here, $\hat{H}_{\text{NL}}(t)=\hbar \Gamma \mathcal{E}(t)\exp{-i\omega_p t}\aid\asd+\ \text{H.c.}$ is the Hamiltonian of a two-mode parametric amplifier excited by monochromatic pump pulses of frequency $\omega_p$, $\ai$ and $\as$ are the annihilation operators for the idler and signal modes, respectively, $\Gamma$ is the nonlinear coupling constant and $\mathcal{E}(t)$ is the temporal envelope of the pump pulses;  $\hat{H}_{\text{JC}}=i \hbar \frac{\Omega}{2}\ai\sgd+\ \text{H.c.}$ is the Jaynes-Cummings (JC) Hamiltonian, where $\Omega$ is the single-photon Rabi frequency and $\sg$ is the atomic lowering operator \cite{jaynes1963JCmodel}; finally, $\hat{H}_0=\hbar(\omega_i\aid\ai+\omega_s\asd\as)+\hbar\omega_0\sgd\sg$ contains the free Hamiltonians of the cavity modes and the emitter. As both the nonlinear and JC interactions are assumed to be resonant (with $\omega_0=\omega_i$ and $\omega_p=\omega_i+\omega_s$), after we transition into the interaction picture via the transformation $\hat{U}_0=\exp{-\frac{i}{\hbar}\hat{H}_0 t}$, the total interaction Hamiltonian takes the form
\begin{equation}\label{S_eq:interactionHamiltonian}
	\hat{H}(t)=\hbar \Gamma \mathcal{E}(t)\aid\asd+i \hbar \Omega\ai\sgd +\ \text{H.c.}
\end{equation}

Below, we attempt to give some more insight into how the introduction of the 2LS impacts the physics of the nonlinear interaction in the hybrid system. We emphasize that we do not see a simple analytical approach for describing and understanding the full hybrid dynamics and, as was done in the main manuscript, we took a numerical approach to obtain the full solution for the dynamics of the system. However, there are physical insights that can be gained about the hybrid dynamics by analytically looking further into the equations of motion of the system. Moreover, by introducing some simplifications and assumptions, one can gain even further insight into how the Rabi oscillation with the 2LS can enhance the probability of generating Fock states in our proposed scheme.

To that end, we first look the state of the hybrid system in the general case, trying to get some analytical insight about the quantum state of the hybrid system in a simple configuration, namely after the first pump pulse, then after Rabi oscillation with the 2LS, and finally after the second pump pulse. With this analysis, we can understand how the Rabi oscillations and the pump pulse for the nonlinear process can generally affect the state of the system. We will then make some strongly simplifying assumptions, to allow us to create a simple and closed-form expression for the state of the system after the second pump pulse, so that we can clearly show how the phases induced by the Rabi oscillation can manipulate the nonlinear interference process and hence result in an enhancement for the probability of generating the $n=1$ state. 

As said, we fist take a look at the general state of the system. We consider the interaction picture and neglect dissipative effects.
The field inside the cavity is initially in the vacuum and ground state $|0,0,g\rangle$. After sending a pulse of gain $r$ into the cavity, a two-mode squeezed vacuum (TMSV) state is generated, where $|\psi\rangle=|\mathrm{TMSV}\rangle=\hat{U}(r)|0,0,g\rangle = \cosh^{-1}(r)\sum_{n=0}^\infty (i\tanh(r))^n|n,n,g\rangle$. Here, $\hat{U}(r)=\mathrm{e}^{-ir(\hat{a}_i\hat{a}_s-\hat{a}_i^\dagger\hat{a}_s^\dagger)}$ is the two-mode squeezing operator \cite{caves1985squeeze}, and we assume a real value for the gain $r$ to keep the present analysis simpler. An important approximation here, which also matches the system parameters in the manuscript, is that the length of the pump pulses is temporally much shorter than the time scale over which the Rabi oscillations take effect. Hence, right after the first pump pulse, we can assume that the state of the system is mainly affected by the squeezing operator, and the effect of the Rabi oscillations is negligible. We emphasize once again, that in our full numerical analysis for obtaining the temporal evolution of the system, we did not make any such approximations. Finally, to make the notation simpler, we write
\begin{equation}\label{S_eq:psi}
|\psi\rangle=|\mathrm{TMSV}\rangle=\sum_{n=0}^\infty \alpha_n(r)|n,n,g\rangle , \   \mathrm{with} \  \alpha_n(r)=\cosh^{-1}(r)(i\tanh(r))^n .
\end{equation}

After the first pump pulse, the interaction continues for a time $T$, during which the generated TMSV state can interact with the 2LS and undergo Rabi oscillations. The interaction with the 2LS creates a series of two-state subspaces, where states $|n,n,g\rangle$ and $|n-1,n,e\rangle$ are coupled to each other and undergo Rabi oscillations with an angular frequency of $\sqrt{n}\Omega$, where $\Omega$ is the single-photon Rabi frequency of the 2LS \cite{scully1997QO}. The resulting state of the system after time $T$ is then
\begin{eqnarray}\label{S_eq:phi}
|\varphi\rangle &=& \alpha_0(r) |0,0,g\rangle +\sum_{n=1}^\infty \alpha_n(r) \left[\cos(\sqrt{n}\Omega T)|n,n,g\rangle+ \sin(\sqrt{n}\Omega T)|n-1,n,e\rangle\right] \nonumber \\
    &=& c_0|0,0,g\rangle+\sum_{n=1}^\infty \left[c_n |n,n,g\rangle + d_n|n-1,n,e\rangle\right].
\end{eqnarray}

As we can see, the main effect of the 2LS is first to create these additional states of the form $|n-1,n,e\rangle$, and also at the same time modulate the amplitude and phase of the different multi-pair states $|n,n,g\rangle$ (as well as the $|n-1,n,e\rangle$ states). This modulation is defined by the factors $\cos(\sqrt{n}\Omega T)$ and $\sin(\sqrt{n}\Omega T)$ and has a $\sqrt{n}$ dependence on the number of photons in the multi-pair component of the state. The real-valued $\cos / \sin$ factors introduce a phase of $0$ or $\pi$ to each multi-pair component (relative to the $\ket{0,0,g}$ state) in a time-dependent fashion, by being either a positive or negative number, respectively. We can also see that if the 2LS was absent (equivalent to setting $\Omega=0$), the state of the system reduces to $|\psi\rangle=|\mathrm{TMSV}\rangle$, where the relative amplitude and phase between the multi-pair components of the state remains unchanged regardless of $T$. 

When a second pulse of gain $r'$ enters the nonlinear cavity, the resulting state of the system becomes
\begin{equation}\label{S_eq:phi_prime}
|\varphi'\rangle= \hat{U}(r') |\varphi\rangle= \alpha_0(r) \hat{U}(r') |0,0,g\rangle +\sum_{n=1}^\infty\alpha_n(r)\left[\cos(\sqrt{n}\Omega T)\hat{U}(r')|n,n,g\rangle +\sin(\sqrt{n}\Omega T)\hat{U}(r')|n-1,n,e\rangle\right].
\end{equation}
We point out that in the absence of interaction with the 2LS, we would have $|\varphi'\rangle= \hat{U}(r') |\psi\rangle= \hat{U}(r')\hat{U}(r)|0,0,g\rangle=  \hat{U}(r+r')|0,0,g\rangle$, where in an example case of pump pulses of equal amplitudes and opposite phases ($r=-r'$) the second pulse will result in a perfect destructive nonlinear interference and take the system's state back to the vacuum state.
Coming back to the general case of Eq.~\eqref{S_eq:phi_prime}, we note that the two-mode squeezing operator can only result in generation or annihilation of photons in pairs. That means $\hat{U}(r')|n,n,g\rangle$ results only in a superposition of states of the form $|m,m,g\rangle$ and $\hat{U}(r')|n-1,n,e\rangle$ results only in a superposition of states of the form $|m-1,m,e\rangle$ ($m\geq1$). Hence, we can formally write out the state after the second pulse as
\begin{eqnarray}  \label{S_eq:phi_prime_1} |\varphi'\rangle&=&c_0'|0,0,g\rangle+\sum_{m=1}^\infty \left[c_m' |m,m,g\rangle + d_m'|m-1,m,e\rangle\right],
\end{eqnarray}
with
\begin{eqnarray}  \label{S_eq:phi_prime_2}
    c_m'&=&\sum_{n=0}^\infty \alpha_n(r) \cos(\sqrt{n}\Omega T) \langle g,m,m|\hat{U}(r')|n,n,g\rangle= \sum_{n=0}^\infty c_n \langle g,m,m|\hat{U}(r')|n,n,g\rangle ,\nonumber\\
    d_m'&=&\sum_{n=1}^\infty \alpha_n(r)\sin(\sqrt{n}\Omega T) \langle e,m,m-1|\hat{U}(r')|n-1,n,e\rangle = \sum_{n=1}^\infty d_n \langle e,m,m-1|\hat{U}(r')|n-1,n,e\rangle.
\end{eqnarray}
By comparing Eqs.(\ref{S_eq:phi_prime_1}) and (\ref{S_eq:phi}) for the state before and after the second pump pulse, we can see that the two states have the same general form, i.e., both being a superposition of $|n,n,g\rangle$ and $|n-1,n,e\rangle$ states. Hence, we can see that the Rabi oscillations and the application of the pump pulse do not affect this general form but each cause a redistribution of the state's coefficients. In particular, Rabi oscillations couple $c_n$ to $d_n$ and vice-versa for each particular $n=1,2,3...$ following the JC model of interaction with a 2LS, whereas each pump pulse causes a rapid and interfering change in the values of $c_n$ and  $d_n$ coefficients following Eq.(\ref{S_eq:phi_prime_2}). The same dynamics will take place for each consecutive Rabi oscillation and incident pump pulse.

Moreover, as can be seen, the coefficients $c_m'$ and $d_m'$, which determine the probability of obtaining a Fock state with $m$ photons in the signal mode ($P_{|m\rangle}=|c_m'|^2+|d_m'|^2$), are determined from an interfering sum of probability amplitudes (the terms $\langle m,m,g|\hat{U}(r')|n,n,g\rangle$ and $\langle m-1,m,e|\hat{U}(r')|n-1,n,e\rangle$ that are responsible for nonlinear interference), weighted by the factors $c_n$ and $d_n$. These themselves contain the factors $\cos(\sqrt{n}\Omega T)$ and $\sin(\sqrt{n}\Omega T)$, resulting in the interaction time $T$ playing a key role in determining the result. 
The generally non-integer factors $\sqrt{n}$ in these weights, introduced by the JC dynamics of the 2LS, radically alter the effects of nonlinear interference induced by the second pump pulse. As was shown with our numerical results in the main manuscript, by applying multiple pump pulses, the individual coefficients $c_m$ and $d_m$ can be controlled to such an extent, that one could maximize these coefficients for individual values of $m$, to allow for generating nearly pure Fock states in the signal mode, when the other degrees of freedom are traced out.

Finally, we can use some strong approximations to be able to derive a short and closed-form expression for the final state of the system after two pulses. In this simplified scheme, we assume that, after the first pump pulse, the generated TMSV state is altered by introducing a relative phase of $\pi$ between the vacuum component of the TMSV (i.e., the $|0,0,g\rangle$ state) and all of the multi-pair states $|n,n,g\rangle$ with $n\geq 1$. We also assume that the probability amplitudes of $|n-1,n,e\rangle$ states are zero. We emphasize that this is not a realistic condition, as the Rabi oscillation introduces factors proportional to sine and cosine for these terms. Yet it is an assumption that allows us to perform a fully analytical calculation to showcase how changing the relative phase between the multi-pair components of a TMSV state can result in a substantially different nonlinear interference once the second pump pulse is incident. 

Under this assumption, we find the state of the system after two pulses. After the first pump pulse of gain $r$, we go from the vacuum state $|0,0,g\rangle$ to a TMSV state $|\psi\rangle=|\mathrm{TMSV} \rangle=\hat{U}(r)|0,0,g\rangle$, as before, which we rewrite now as $|\psi\rangle=\cosh^{-1}(r)|0,0,g\rangle+|\phi\rangle$ \cite{caves1985squeeze}, where $|\phi\rangle =\cosh^{-1}(r)\sum_{n=1}^\infty (i\tanh(r))^n|n,n,g\rangle$ is the sum of all multi-pair components of the state. Now we apply the "phase flip" to the multi-pair components and obtain $|\psi'\rangle=\cosh^{-1}(r)|0,0,g\rangle-|\phi\rangle=2\cosh^{-1}(r)|0,0,g\rangle-|\psi\rangle$. If we now send a second pulse of gain $-r$ into the cavity, we obtain $|\psi''\rangle=2\cosh^{-1}(r)\hat{U}(-r)|0,0,g\rangle-\hat{U}(-r)|\psi\rangle$. By noting that $\hat{U}(-r)|0,0,g\rangle$ is again a TMSV state, and that $\hat{U}(-r)|\psi\rangle=\hat{U}(-r)\hat{U}(r)|0,0,g\rangle=|0,0,g\rangle$ \cite{caves1985squeeze,chekhova2016NonlInterf}, we can write out the first two terms of the final state $$|\psi''\rangle=(2\cosh^{-2}(r)-1)|0,0,g\rangle+2i\cosh^{-2}(r)\tanh(r)|1,1,g\rangle+... \: ,$$
where we see that the probability of detecting a single pair state $|1,1,g\rangle$, and consequently, a single photon in each mode, is now $4\cosh^{-4}(r)\tanh^2(r)$. This probability maximizes for a parametric gain of $r\approx5.7\text{dB}$ (equivalently $r\approx 0.66$) and attains the maximum probability value of $P_{|1\rangle}\approx0.59$, more than twice the maximum value attainable using only nonlinear interactions. We also note that although we made quite strong approximations for analytically finding this result with the two-pulse case, our finding is also not too far off the numerical results shown in Fig.2 and Fig.1(e). Thus, just by altering the relative phase between the multi-pair components of the TMSV in a particular way, the nonlinear interference mechanism between the photon pairs generated by the two pump pulses can be altered and can be made to result in a state with a significantly enhanced Fock state component.

\section{Effects of pump relative phase in a two-pulse configuration}

\begin{figure*}[htpb!]
	\includegraphics[width=0.9\linewidth]{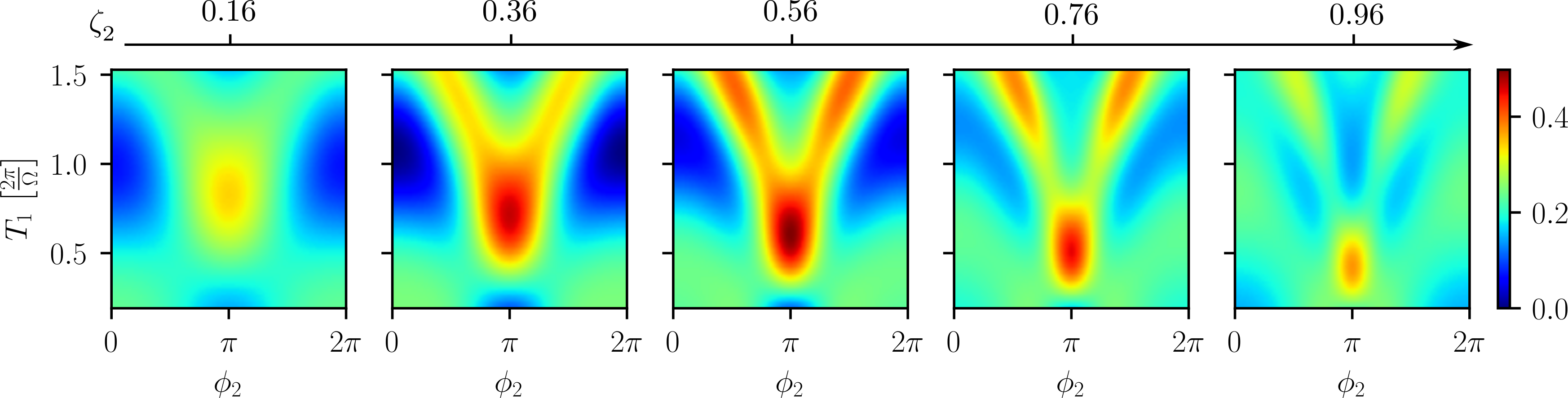}
	\caption{The probability of having a single photon in the signal mode in a two pulse configuration, as a function of the parametric gain phase $\phi_2$ and delay time $T_1$ of the second pulse. The second pulse has a fixed gain magnitude $\zeta_2$ for each plot, with the value increasing from left to right, as indicated by the axis above the figures. In all cases the gain of the first pulse is fixed to $r_1\approx0.58$.}
	\label{fig_s1}
\end{figure*}

As discussed in the main text, the relative phase between the pump pulses exciting the nonlinear cavity is one of the parameters that influences the final distribution of multi-pair states at the output of the hybrid source. In the two-pulse example in Fig.~2(b), we showed the dependence of the probability of obtaining a single photon in the signal mode as a function of the magnitude $\zeta_2$ and phase $\phi_2$ of the parametric gain of the second pump pulse, for a fixed delay time $T_1$. For the set of parameters and parameter ranges investigated, the single photon probability exhibits a maximum for $\phi_2=\pi$. While our simulations indicate (as will be shown here) that a \textbf{global} maximum for the three parameters considered $\{\zeta_2,\phi_2,T_1\}$ is always attained for $\phi_2=\pi$, the single photon probability can exhibit local maxima at different values of $\phi_2$, when $T_1$ is varied. This behavior is shown in Fig.~\ref{fig_s1}, where we see that the single photon probability (again calculated by adding the probabilities of the $\ket{1,1,g}$ and $\ket{0,1,e}$ states) can exhibit either a single maximum for $\phi_2=\pi$ at shorter $T_1$, or multiple maxima at $\phi_2\neq\pi$ at longer $T_1$, with the exact value of $T_1$ at which the behavior changes depending on the value of $\zeta_2$. Regardless of $\zeta_2$, the single maximum at $\phi_2=\pi$ is always the highest value.

\section{Decoupled dynamics and numerical procedures}
In order to efficiently perform a sweep of parameters, to evaluate the possible output states for the three-pulse configuration discussed in the main text, we make use of the assumption that the duration of each pump pulse is much shorter than the few-photon Rabi periods of the 2LS. As we will show here, this allows the nonlinear and 2LS dynamics to be evaluated separately and avoids the need for integrating the full Schrödinger equation for each set of pulse parameters.

In the interaction picture, the evolution operator of the hybrid system is written as
\begin{equation}\label{S_eq:fullEvolutionOperator}
    \hat{U}(t)=\mathcal{T}\mathrm{exp}\left[-\frac{i}{\hbar}\int_{-\infty}^t\dd{t'}\hat{H}(t')\right],
\end{equation}
where, $\hat{H}(t')$ is given by Eq.~\eqref{S_eq:interactionHamiltonian} and $\mathcal{T}$ indicates the time-ordering superoperator. By expanding the interaction Hamiltonian according to Eq.~\eqref{S_eq:interactionHamiltonian} we obtain
\begin{equation}\label{S_eq:expandedEvolutionOperator}
    \hat{U}(t)=\mathcal{T}\mathrm{exp}\left\{-\frac{i}{\hbar}\int_{-\infty}^t\dd{t'}\left[\hbar \Gamma \mathcal{E}(t')\aid\asd+i \hbar \Omega\ai\sgd +\ \text{H.c.}\right]\right\}.
\end{equation}
We now assume that the pump field envelope $\mathcal{E}(t')$ consists of a sequence of temporally distinct pulses as $\mathcal{E}(t')=\sum_j\mathcal{E}_j(t')$, where $\mathcal{E}_j(t')=\abs{\mathcal{E}_j(t')}\exp{i\phi_j}$ is the envelope of the $j$-th pulse consisting of a time-dependent amplitude $\abs{\mathcal{E}_j(t')}$ and arbitrary \textit{time-independent} phase $\phi_j$. 

We can divide the total interaction time into intervals during a pump pulse, of duration $\delta T_j$, and intervals in-between pulses, of duration $T_j$. In-between the $j$-th and $j+1$-th pulses, we have $\mathcal{E}(t)\approx0$ and can neglect the nonlinear interaction, thus, the evolution over that time interval is well approximated by 
\begin{equation}\label{S_eq:2lsEvolution}
\hat{U}^{(j)}_\mathrm{2LS}=\mathrm{exp}\left[T_j \Omega \ai\sgd - \mathrm{H.c.} \right].
\end{equation}
For intervals during a pulse, the assumption $\delta T_j\ll \frac{2\pi}{\Omega}$ results in the contributions from the 2LS interaction being negligible in comparison to the nonlinear interaction. Thus, the evolution of the system due to the $j$-th pump pulse is well-approximated by $\hat{U}^{(j)}_\mathrm{NL}=\mathcal{T}\mathrm{exp}\left\{-i\Gamma\int_j\dd{t'} \left[\mathcal{E}_j(t')\aid\asd+\mathcal{E}_j^*(t')\ai\as\right]\right\}$. The form of $\hat{U}^{(j)}_\mathrm{NL}$ can be further simplified by recalling $\mathcal{E}_j(t')=\abs{\mathcal{E}_j(t')}\exp{i\phi_j}$ and rewriting the integrand in the exponential as $\abs{\mathcal{E}_j(t')}\left(\exp{i\phi_j}\aid\asd+\exp{-i\phi_j}\ai\as\right)$, which leaves the operator part of the integrand \textit{time-independent} and eliminates the need for explicit time-ordering in the expression for $\hat{U}^{(j)}_\mathrm{NL}$, which takes on the form
\begin{equation}\label{S_eq:nlEvolution}
    \hat{U}^{(j)}_\mathrm{NL}=\mathrm{exp}\left[-i \left(r_j\aid\asd+r^*_j\ai\as\right)\right],
\end{equation}
where we defined $r_j=\Gamma\int_j\dd{t'}\mathcal{E}_j(t')$. This operator has the exact form of a two-mode squeezing operator with the parametric gain $r_j=\zeta_j\exp{i\phi_j}$ \cite{caves1985squeeze} and generates a TMSV state when acting on a vacuum initial state. Another consequence of the definition Eq.~\eqref{S_eq:nlEvolution} is that the nonlinear part of the evolution is completely described by the parametric gain associated with the pulse and the exact shape of the pulse does not affect the result.

By taking into account the above conclusions, the total evolution operator of the hybrid system up to the time $t$, excited by a series of temporally-distinct pulses can be represented as
\begin{equation}\label{S_eq:decomposedEvolution}
    \hat{U}(t)\approx\prod_j \hat{U}^{(j)}_\mathrm{2LS}\hat{U}^{(j)}_\mathrm{NL},
\end{equation}
where the pairs of operators are ordered from left to right in descending order with respect to $j$. 

In order to efficiently evaluate the state at the output of a large set of pump pulse configurations, we calculated the matrix representation of $\hat{U}^{(j)}_\mathrm{2LS}$ and $\hat{U}^{(j)}_\mathrm{NL}$ in a sufficiently large basis set and sequentially applied them to the initial state $\ket{0,0,g}$. The process was repeated for all combinations of pulse parameters involved, i.e. $\{\zeta_1,T_1,\zeta_2,T_2,\zeta_3\}$, to obtain the various parameter dependencies shown in Figs.~2 and \ref{fig_s1}, as well as to find the optimal configurations for the particular Fock states, whose temporal evolution is shown in Fig.~3. In Table~\ref{table_S:pulseParams}, we show the optimal three-pulse-configuration parameters, corresponding to the maximum probabilities for obtaining the Fock states with up to $n_s=9$, shown in Fig.~3(d).

Once the optimum pulse configurations were found using the above-explained method, we obtained and plotted the temporal evolution of the system for particular pulse configurations. This was done using the \texttt{QuTiP} library in Python \cite{johansson2012qutip}, both to solve the Schrödinger equation for the lossless system (shown in Figs.~1 and 3) and to solve the Lindblad master equation in the lossy system. 

Both for sweeping over pump pulse parameters, as well as obtaining the temporal evolution of the system for particular configurations, we found that a composite basis $\left\{\ket{n_i,n_s,g \text{ or } e};\ n_{i,s}=0,1,... 59\right\}$ with a photon basis size of $60$ states per mode represents a good balance between numerical accuracy and computational resource requirements. Yet, we note that this number depends on the total parametric gain that the system experiences and a larger composite basis should be used if higher gain magnitudes or more pulses are considered.

\begin{figure*}[htpb!]
	\includegraphics{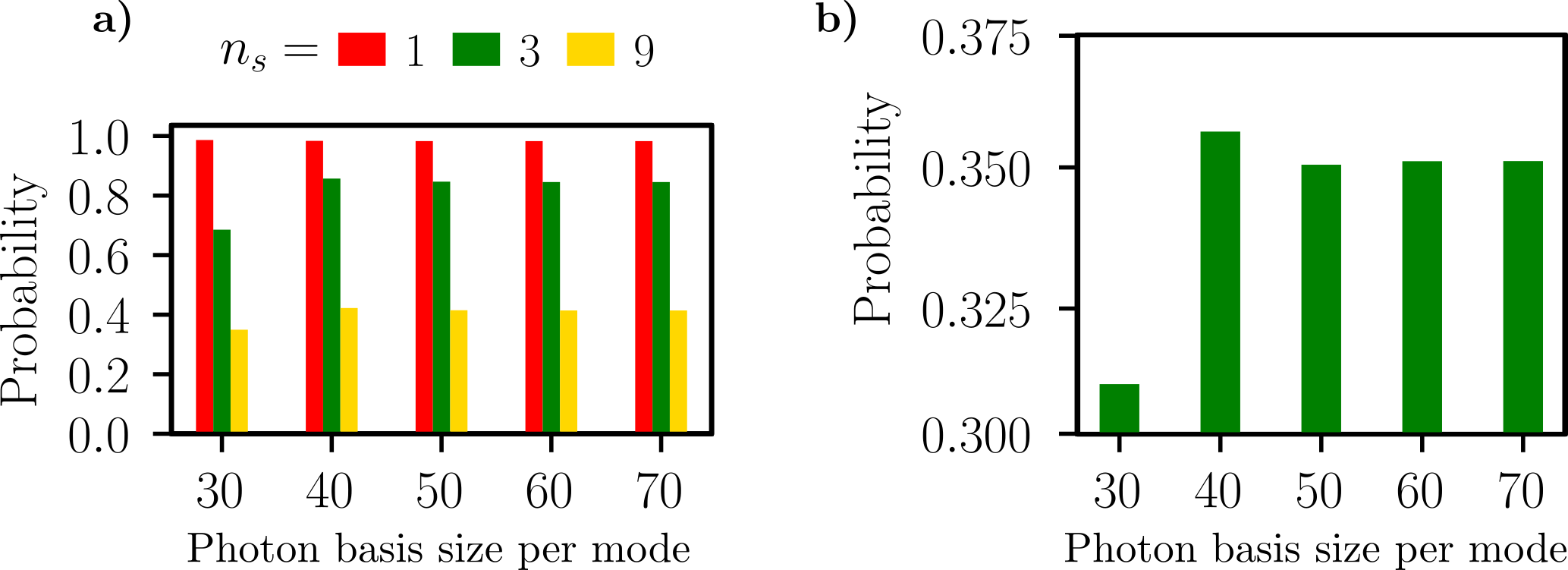}
	\caption{\textbf{(a)} Output probabilities of detecting $1$,$3$ and $9$ photons in the signal mode, each corresponding to the optimal pulse configuration from Table~\ref{table_S:pulseParams} and without loss and dephasing, calculated for different basis sizes. \textbf{(b)} The probability of detecting a $3$-photon Fock state in the signal mode using the optimal pulse configuration and in the presence of cavity loss with $\gamma_c=0.03\Omega$, calculated for different basis sizes.}
	\label{fig_s2}
\end{figure*}

In Fig.~\ref{fig_s2} we show the dependence of numerically calculated probabilities on the basis size. The output probabilities of detecting $1$, $3$ and $9$ photons in the signal mode, each obtained using the optimal pulse configurations from Table~\ref{table_S:pulseParams} (considering the lossless and no dephasing case) and calculated for different basis sizes are shown in Fig.~\ref{fig_s2}(a). Similarly, in Fig.~\ref{fig_s2}(b), we show the probabilities for obtaining a $3$-photon state in the signal mode in the presence of cavity decay with $\gamma_c=0.03\Omega$, also as a function of the basis size. All of the values converge to within $\sim10^{-3}$ for a photon basis size of $60$ in each cavity mode.

\begin{table}[htpb]                                                          
\centering                                                             
\begin{tabular}{|c|c|c|c|c|c|c|}                                                                                                      
\hline                                                                                                                                
\rule{0pt}{3ex}Fock state & $\mathcal{P}$ & $\zeta_1$ [dB] & $T_1$ [$\frac{2\pi}{\Omega}$] & $\zeta_2$ [dB] & $T_2$ [$\frac{2\pi}{\Omega}$] & $\zeta_3$ [dB] \\[3pt]
\hline                                                                                                                                
$\ket{1}$ & 0.98 & 4.76 & 1.11 & 12.86 & 0.19 & 12.39 \\                                                                              
\hline                                                                                                                                
$\ket{2}$ & 0.93 & 8.10 & 1.41 & 12.86 & 0.34 & 10.96 \\                                                                              
\hline                                                                                                                                
$\ket{3}$ & 0.85 & 9.53 & 1.49 & 13.34 & 0.50 & 10.00 \\                                                                              
\hline                                                                                                                                
$\ket{4}$ & 0.74 & 9.53 & 1.49 & 13.34 & 0.65 & 9.53 \\                                                                               
\hline                                                                                                                                
$\ket{5}$ & 0.65 & 8.57 & 1.49 & 13.34 & 0.80 & 9.05 \\                                                                               
\hline                                                                                                                                
$\ket{6}$ & 0.58 & 8.57 & 1.49 & 13.34 & 0.95 & 8.57 \\                                                                               
\hline                                                                                                                                
$\ket{7}$ & 0.52 & 9.05 & 1.49 & 13.34 & 1.11 & 8.10 \\                                                                               
\hline                                                                                                                                
$\ket{8}$ & 0.47 & 9.53 & 1.49 & 13.34 & 1.26 & 7.62 \\                                                                               
\hline                                                                                                                                
$\ket{9}$ & 0.42 & 10.00 & 1.49 & 13.34 & 1.41 & 7.15 \\                                                                              
\hline                                                                                                                                
\end{tabular}                                                           
\caption{The maximum probabilities and corresponding parameter values for the optimal pump pulse configurations for generating Fock states up to $n_s=9$ in the three-pulse configuration.}                                               
\label{table_S:pulseParams}                                             
\end{table}        

\section{Estimation of Purcell factor from cavity decay rate for realistic optical systems}\label{sec:Purcell}
We assume a spontaneous decay rate of $\Gamma_0=10^9s^{-1}$ for the 2LS, which is in the same order of magnitude of many common solid-state quantum emitters \cite{Lodahl2015, he2015single, lee2020integrated}.
With a center wavelength of the idler mode at $\lambda_0=600$nm, and a mode volume of $V=10\lambda_0^3$ for the cavity, we find the single-photon Rabi frequency $\Omega=\sqrt{\frac{3 c \Gamma_0 \lambda_0^2}{2\pi V}}\approx1.6\times10^{11}$ $s^{-1}$ \cite{scully1997QO}, and can use it to calculate the quality factors $Q=\frac{2\pi c}{\lambda_0\gamma_c}$, associated with each of the values of $\gamma_c$. Thus, to achieve system dynamics corresponding to the solid lines in Fig.~4(a,b,c), we require a cavity with a quality factor of $Q\approx1.9\times10^7$, while to obtain dynamics corresponding to the dashed and dotted lines, we require quality factors of $Q\approx2.5\times10^6$ and $Q\approx6.5\times10^5$, respectively.
Although the range of investigated values of $\gamma_c$ correspond to the strong coupling regime ($\gamma_c \ll \Omega$), we can calculate a Purcell factor for the cavity $F=\frac{3}{4\pi^2}\frac{Q}{V/\lambda^3}$ \cite{Lodahl2015}, corresponding to these three $Q$ values. We then find the values of $F\approx 5\times 10^3$, $F\approx 1.9\times 10^4$, and $F\approx 1.4\times 10^5$ for the smallest to largest of the quality factors. We point out, that although these Purcell factors are quite large, they are within reach in realistic photonic structures, such as photonic crystal cavities \cite{deotare2009high}, fiber-based microcavities \cite{PhysRevX.10.041025}, or hybrid-plasmonic cavities \cite{PhysRevApplied.18.044066}.

\section{Potential technological platforms for realizing the proposed hybrid system}

For realizing the envisioned hybrid system, we require three fundamental components: a two-mode cavity, nonlinear material, and a two-level system (2LS), all operating around the optical domain (if the target operation wavelengths are in the optical domain). It is important to note that the nonlinear material together with the two-mode cavity has the function of generating the two-mode squeezed vacuum state (TMSV). This can be done using materials with both second-order $\chi^{(2)}$ and third-order $\chi^{(3)}$  nonlinearity. Moreover, we do not necessarily need a piece of nonlinear material that is placed inside a bulk cavity (as shown in our schematic Fig.1(a) in the main manuscript), although such systems have been the workhorse of the squeezed-light generation community \cite{schnabel2016squeeze}. This schematic more generally represents a nonlinear cavity, which can also be realized by nanostructuring a nonlinear material platform for creating on-chip photonic elements like microring resonators or photonic crystal cavities. In fact, even triply resonant nonlinear interactions (a resonance at each of the pump, signal, and idler frequencies) can be realized with nanostructured nonlinear resonators (nanobeam cavity on the silicon platform using the $\chi^{(3)}$-nonlinearity: \cite{azzini2013stimulated}; photonic crystal cavity on the InGaP platform using its $\chi^{(3)}$-nonlinearity: \cite{marty2021photonic}; microring resonator on the lithium niobate platform using its $\chi^{(2)}$ : \cite{lu2021ultralow}). The frequency of these resonances can be controlled by the geometric parameters of the nanostructured cavity design. Finally, for the 2LS component, any solid-state single-photon emitter, such as quantum dots (QDs) or defect centers, is generally a viable candidate for our scheme, as it can effectively act as a 2LS.

With all this said, for the realization of our proposed hybrid system, we require a nonlinear material platform with strong $\chi^{(2)}$  or $\chi^{(3)}$  nonlinear coefficient, with developed nanostructuring technology for creation of high quality on-chip resonator elements, which at the same time can also host a solid-state single-photon emitter. There are in fact several such material platforms at hand. One example is the material platform of GaAs (combined with other compounds like AlGaAs, InAs, AlAs that can be grown heterogeneously with it), which has an advanced nanostructuring technology, has strong $\chi^{(2)}$  and $\chi^{(3)}$  nonlinearity used for realizing nanostructured sources of photon pairs \cite{steiner2021ultrabright,orieux2013direct}, and can host very high-quality QD single-photon emitters \cite{senellart2017high,ding2016demand}. Another material platform is silicon and silicon nitride, which also have an advanced nanostructuring technology, have strong $\chi^{(3)}$  nonlinearities, used for realizing nanostructured sources of photon pairs and squeezed light \cite{barone2024generation,zhang2021squeezed}, and can also host defect centers \cite{johnston2024cavity,senichev2021room}. Another interesting platform, which we find very promising for realization of our proposed hybrid system, is silicon carbide (SiC), which is a material with a strong $\chi^{(2)}$-nonlinearity, with very recent demonstrations of pair generation in a microring resonator \cite{li2024integrated}, and can also host vacancy centers for single-photon emission \cite{castelletto2021silicon}. What we find particularly interesting about SiC, is that it combines the above features with a wide bandgap property, which could be useful for avoiding unwanted fluorescence and two-photon absorption, and also its wide transparency window allows a wider range of possibilities for frequencies of operation.
We should mention that there is generally also the possibility of mixing and matching different material platforms, through methods like wafer bonding, transfer printing, and pick and place methods \cite{elshaari2020hybrid}), to get access to even a larger set of possibilities for creating our proposed hybrid system. For example, this has been done with an InAs QD that was in a InP nanobeam, where the nanobeam was moved with a microprobe tip and very accurately placed on top of a nanostructured thin film lithium niobate nanostructured ridge waveguide \cite{aghaeimeibodi2018integration}. Such a combination could be very interesting for our hybrid system, where one could essentially combine the well-developed capability of lithium niobate for squeezed-light generation and optical parametric oscillation \cite{park2024single,lu2021ultralow,nehra2022few} with the highly advanced performance of semiconductor QDs as effective two-level systems \cite{senellart2017high}. Moreover, flakes of 2D materials with embedded single-photon emitters have also been placed accurately using dry transfer methods onto nanostructured components \cite{peyskens2019integration}. Although such hybridization approaches are generally more technologically challenging and possibly not suited for large-scale fabrication methods, they are already demonstrated and can create countless opportunities for creating our proposed hybrid system.
As discussed above, there are quite many technological platforms and approaches that are available for realizing our envisioned hybrid system. Which final platform one chooses, in the end depends on many factors that should all come together in an optimized design approach that also considers practical aspects of the implementation. This include (but are not limited to): what frequency range is desired for the generated signal photons, what transparency window is needed for the nonlinear material, what is the range of transition frequencies for the single-photon emitters that can be embedded or combined with that cavity material platform, what nonlinear process we rather use (with $\chi^{(2)}$  or $\chi^{(3)}$ ), and so on. In the end, it could be that unforeseen technological or fundamental challenges arise for operating such hybrid nonlinear systems in practice, which were not encountered in the stand-alone operation regime of a squeezed-light source and a single-photon source, but these will become clearer with proof-of-principle experimental realizations of such hybrid systems.

\section{Hybrid dynamics in the presence of 2LS dephasing}
To quantify the effects of dephasing on the performance of the hybrid scheme, we solved the Lindblad master equation for different values of the dephasing rate $\gamma_d=\{0.001,0.005,0.01,0.005\}\Omega$, while neglecting cavity losses to better isolate the influence of dephasing. For a 2LS with properties as given in Sec.~\ref{sec:Purcell}, these values of $\gamma_d$ correspond to dephasing times in the range $125$ps-$6.25$ns, which includes commonly measured values for many types of QD’s realized in modern solid-state platforms \cite{press2010ultrafast,dietrich2016gaas,esmann2024solid}. 

\begin{figure*}[htpb!]
	\includegraphics[width=\linewidth]{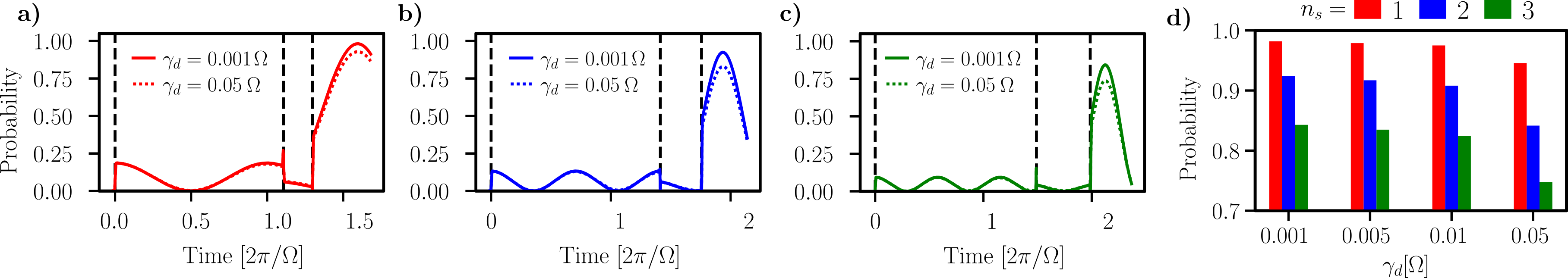}
	\caption{\textbf{(a-c)} Temporal evolution of the $\ket{1,1,g}$ (red), $\ket{2,2,g}$ (blue), and $\ket{3,3,g}$ (green) states in the presence of 2LS dephasing (for three different pulse configuration parameters), with rates quantified by $\gamma_d$. Since the effects of dephasing only become sufficiently relevant for the highest rate considered, only the dynamics corresponding to the lowest and highest rates are shown in the plots. The pump configuration for each of the plots is identical to the corresponding optimized lossless cases shown in Table~\ref{table_S:pulseParams} and also shown in Figs.3(a-c) in the main manuscript; \textbf{(d)} The probabilities of detecting $n_s=1,2,3$ photons in the signal mode immediately after the third pump pulse for the three different pulse configurations, shown for different values of $\gamma_d$. To make the differences between results obtained for different $\gamma_d$ more discernible, the range of the y-axis has been adjusted appropriately.}
	\label{fig_s3}
\end{figure*}

We study three cases, where the system parameters correspond to the cases shown in Figs.~3(a-c) in the main manuscript, where the system was optimized for generation of $1$-, $2$- and $3$-photon states in the signal mode. The obtained dynamics of the hybrid system in the presence of dephasing are shown in Fig.~\ref{fig_s3}(a-c), for two different values of dephasing rate for clearer visuals. The probabilities of detecting $n_s=1,2,3$ photons in the signal mode immediately after the third pump pulse, for different values of d, are shown in Fig.~\ref{fig_s3}(d). Comparing to the results in the presence of cavity losses (shown in Fig.~4 of the main manuscript), we observe that pure 2LS dephasing has a considerably weaker detrimental effect on the optimized output probabilities of the hybrid system, only becoming prominent for $\gamma_d=0.05\Omega$ (i.e., a dephasing time of $125$ps). Hence, we can conclude that photon losses are a much more detrimental effect than the 2LS dephasing. 

We note here that similar to our loss investigation in the main manuscript, we investigated dephasing also with configurations that were optimized under a no-dephasing condition. Hence, detrimental effects from dephasing and losses could potentially be reduced by performing the optimization for generation of Fock states in the presence of dephasing and loss, and also potentially by including more free parameters for the system.

\bibliography{bibliography}